\documentclass[pre,
 amsmath,amssymb,superscriptaddress,aps,notitlepage
]{revtex4-1}

\usepackage{graphicx}
\usepackage[toc,page]{appendix}
\usepackage[normalem]{ulem}
\usepackage[table]{xcolor}
\usepackage{dcolumn}
\usepackage{bm}
\usepackage{float}
\usepackage{braket}

\newcommand{\beq}{\begin{equation}}
\newcommand{\eeq}{\end{equation}}

\begin{document}

\title{Electrostatic complementarity at the interface drives transient protein-protein interactions}

\author{Greta Grassmann}
\affiliation{Department of Biochemical Sciences ``Alessandro Rossi Fanelli", Sapienza University of Rome, P.Le A. Moro 5, 00185, Rome, Italy}

\author{Lorenzo Di Rienzo}

\affiliation{Center for Life Nano \& Neuro Science, Istituto Italiano di Tecnologia, Viale Regina Elena 291,  00161, Rome, Italy}

\author{Giorgio Gosti}

\affiliation{Center for Life Nano \& Neuro Science, Istituto Italiano di Tecnologia, Viale Regina Elena 291,  00161, Rome, Italy}

\affiliation{Soft and Living Matter Laboratory, Institute of Nanotechnology, Consiglio Nazionale delle Ricerche, 00185, Rome, Italy}

\author{Marco Leonetti}

\affiliation{Center for Life Nano \& Neuro Science, Istituto Italiano di Tecnologia, Viale Regina Elena 291,  00161, Rome, Italy}

\affiliation{Soft and Living Matter Laboratory, Institute of Nanotechnology, Consiglio Nazionale delle Ricerche, 00185, Rome, Italy}

\author{Giancarlo Ruocco}

\affiliation{Center for Life Nano \& Neuro Science, Istituto Italiano di Tecnologia, Viale Regina Elena 291,  00161, Rome, Italy}

\affiliation{Department of Physics, Sapienza University, Piazzale Aldo Moro 5, 00185, Rome, Italy}

\author{Mattia Miotto \footnote{The authors contributed equally to the present work.} \footnote{Corresponding authors:\\ mattia.miotto@roma1.infn.it\\ edoardo.milanetti@uniroma1.it}}

\affiliation{Center for Life Nano \& Neuro Science, Istituto Italiano di Tecnologia, Viale Regina Elena 291,  00161, Rome, Italy}

\author{Edoardo Milanetti \footnotemark[1]\footnotemark[2] }
\affiliation{Department of Physics, Sapienza University, Piazzale Aldo Moro 5, 00185, Rome, Italy}

\affiliation{Center for Life Nano \& Neuro Science, Istituto Italiano di Tecnologia, Viale Regina Elena 291,  00161, Rome, Italy}

\begin{abstract}

Understanding the molecular mechanisms driving the binding between bio-molecules is a crucial challenge in molecular biology. 
In particular, the comprehension of the recognition/docking determinants, together with those of complex formation and stability have deep implications 
for the prediction of the binding region, which constitutes the essential starting point for drug-ability and design.
Indeed, physicochemical features of the binding regions have been subject to intense investigations to identify binding hallmarks to be used by predictive tools.   
In this respect, characteristics like the preferentially hydrophobic composition of the binding interfaces, the role of van der Waals interactions (short range forces), and the consequent shape complementarity between the interacting molecular surfaces are well established. However, no consensus has yet been reached on how and how much electrostatic participates in the various stages of protein-protein interactions, i.e. far recognition of binding partners, binding adaptation, and proper binding stability. 
Here, we perform extensive analyses on a large dataset of protein complexes for which both experimental binding affinity and pH data were available.
Probing the amino acid composition, the disposition of the charges, and the electrostatic potential they generated on the protein molecular surfaces, we found that (i) although different classes of dimers do not present marked differences in the amino acid composition and charges disposition in the binding region, (ii)  homodimers with identical binding region show higher electrostatic compatibility with respect to both homodimers with non-identical binding region and heterodimers. The level of electrostatic compatibility also varies with the pH of the complex, reaching the lowest values for low pH. The opposite behaviors can be observed for shape complementarity, whose values are highest when the pH is low.
Interestingly, (iii) shape and electrostatic complementarity, for patches defined on short-range interactions, behave oppositely when one stratifies the complexes by their binding affinity: complexes with higher binding affinity present high values of shape complementarity (the role of the Lennard-Jones potential predominates) while electrostatic tends to be randomly distributed. Conversely, complexes with low values of binding affinity exploit Coulombic complementarity to acquire specificity, suggesting that electrostatic complementarity may play a greater role in transient (or less stable) complexes. In light of these results, (iv) we provide a fast and efficient method to measure electrostatic complementarity without the need of knowing the complex structure. Expanding the electrostatic potential on a basis of 2D orthogonal polynomials, we can discriminate between transient and permanent protein complexes with an AUC of the ROC of $\sim$ 0.8.
\end{abstract}


\maketitle

\section{Introduction}

Interactions among proteins constitute the molecular basis of most processes in living organisms, and their deregulation or disruption often leads to disease \cite{Ryan2005-uz, Rabbani2018-lf, Gracia2022}. Among other things, such interactions may differ in the number (dimers, tetrameters, etc) and kind of involved proteins (homo or hetero complexes), the stability of the binding (transient/permanent bindings), and the type of the binding process, i.e. lock and key, induced fit and conformational selection. 
While it has been estimated that over 80\% of proteins operate in molecular complexes \cite{https://doi.org/10.1002/pmic.200700131}, detailed comprehension of the mechanism behind the protein binding process and the stability of the resulting protein complexes is still incomplete.
At a qualitative level, binding involves a recognition phase where distant molecules have to recognize themselves in the crowded cellular environment followed by a docking process where the two molecules reorient/adapt to binding in specific regions.
Despite this, complex formation is often highly specific: a binding partner could be recognized by only one of the members in a protein family even if they all have the same folds \cite{Sheinerman2002}. This compatibility is determined by an interplay between various contributions on the molecular surface and can either (i) be present from the beginning, when the two proteins are far apart (lock and key model), or (ii) be assumed by the proteins while exploring their conformational landscape (conformational selection model) or be gained while interacting with the partner (induced fit model) \cite{koshland1995key,csermely2010induced,paul2016distinguish}. However, once the proteins are bound, their binding regions are known to display a combination of geometrical and chemical complementarities, which ultimately reflect on the binding stability \cite{Gabb1997, Desantis2022, skrabanek2008computational, van1994protein, Miotto_2018}.

At the level of amino acid composition, it is widely known that the composition of binding regions is different with respect to the rest of the solvent-exposed region: while the latter is preferentially populated by hydrophilic residues, binding regions have a higher number of hydrophobic residues, like Val and Leu, that tend to establish stronger van der Waals interactions~\cite{Desantis2022, erijman2014structure}.
From a geometrical point of view, the optimization of short-ranged interactions between atoms at the interface leads to a local shape complementarity of the proteins' molecular surfaces. 
Indeed, the side chain rearrangements minimizing the van der Waals interaction determine shape complementarity at the interfaces, which is typically evaluated by geometrical approaches~\cite{Milanetti2021, kihara2011molecular,gainza2020deciphering, daberdaku2019antibody,zhu2015large, venkatraman2009protein}.

Conversely, there is still no full consensus on the role played by electrostatic interactions, including hydrogen bonding, ionic/Coulombic, cation$-$ $\pi$, $\pi-\pi$, lone-pair sigma hole, and orthogonal multipolar interactions~\cite{Bauer2019, Sheinerman2002, Di_Rienzo2021-yp, 10.3389/fmolb.2021.626837, MILANETTI2019360351}. In fact, acting at longer distances, it is unanimously understood that electrostatic compatibility plays a role at the beginning of the recognition process when partners are far away from each other \cite{Shashikala2019}; in fact, proteins move in a very crowded environment and since electrostatic interactions are the most long-ranged ones, they produce a drift in the Brownian motion of the two binding proteins. However, while this could be true for heterodimers (that may possess opposite charges), homodimers have the same net charge, thus attractive interactions can only take place between parts of the proteins at the most~\cite{Zhang2011}. 
Therefore, many studies are focusing on assessing the electrostatic match of protein complexes, to better understand why and how binding happens \cite{Vascon2020, Zhang2011, Kundrotas2006, Tsuchiya2006, Zhou2018, Gabb1997, Yoshida2019}.

In particular, McCoy \textit{et al.}~\cite{McCoy1997} found that binding sites are characterized by significant values of this quantity defining electrostatic complementarity as the correlation of surface electrostatic potential at binding sites on a small number of protein complexes.
Another study discussed the importance of electrostatic interactions in the binding adaptation \cite{Ghaemi2017}.
Shashikala and coworkers~\cite{Shashikala2019} investigated the role of electrostatic interactions in diseases, finding that disease-causing mutations frequently alter wild-type electrostatic interactions.
Moreover, electrostatic turned out to be a key feature even for machine learning methods that look at the identification of protein-protein binding sites  \cite{gainza2020deciphering}.
Similarly, electrostatic and shape complementarity turned out to be sufficient to predict the DNA-binding sites on proteins with 80\%  accuracy \cite{Tsuchiya2006}. 

\begin{figure}[] 
\begin{center}
\includegraphics[width=0.8\textwidth]{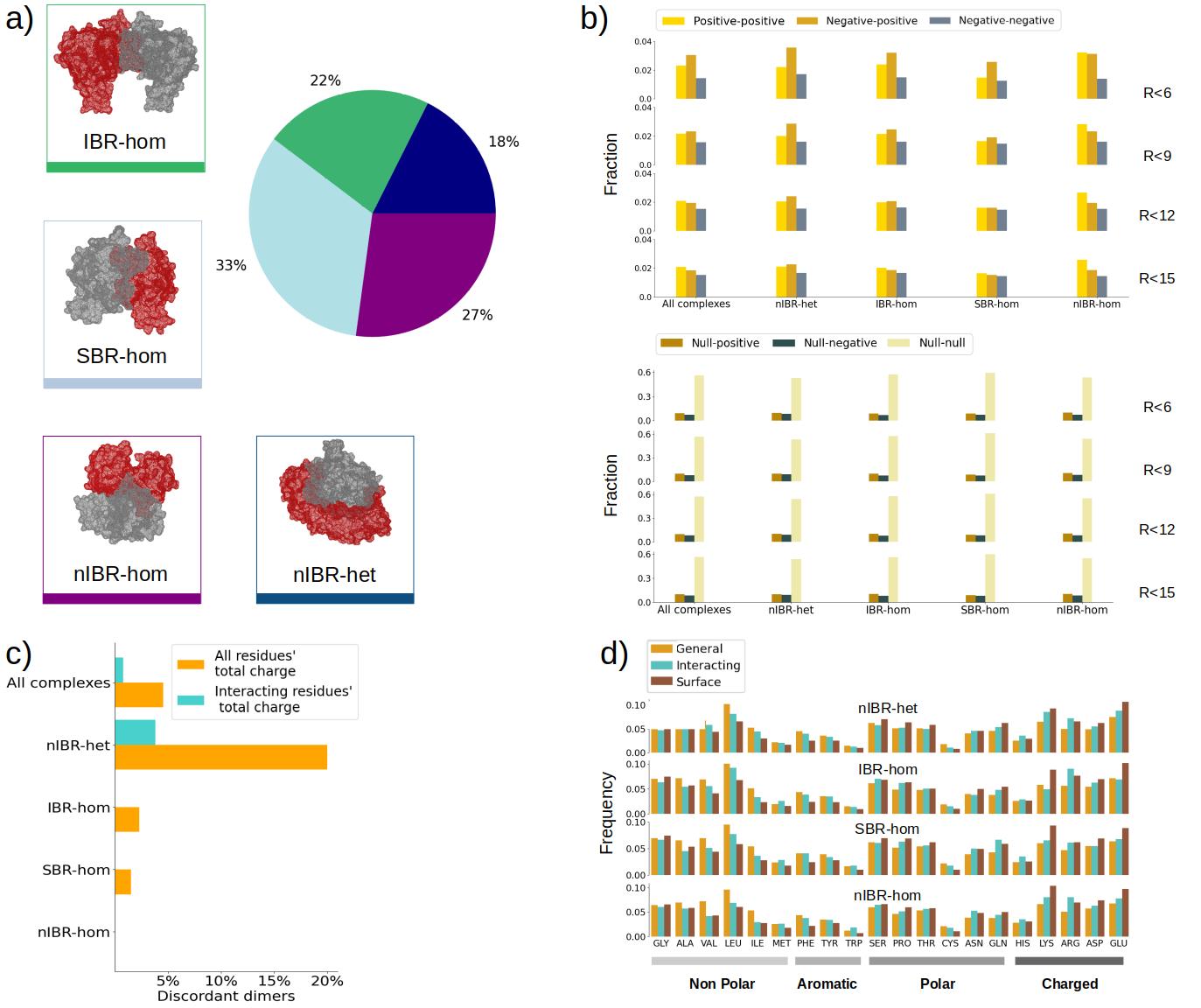}
\caption{\textbf{Amino acid composition, charge properties, and classification of the dataset.} \textbf{a)} The complexes in the dataset are divided into heterodimers and IBR, SBR and nIBR homodimers. The colored boxes report an example for each category. The same colors are used to indicate in the pie chart each class abundance in the dataset. \textbf{b)} On each interacting surface residue a sphere of radius R is built, and the number of interacting residues on the partner's surface included in the sphere is counted.
The top bar plot shows, for increasing values of R (as reported by the labels on the right) and for both the whole dataset and each of the four classes, the fraction of positively or negatively charged residues that can be found close to positive residues, respectively in yellow or ochre. In grey, the fraction of negative residues closed to a negative amino acid.
The bottom bar plot originates from the same analysis: the bars in brown, dark blue, and cream show respectively the fraction of positively, negatively, and null charged residues close to the non-charged ones.
\textbf{c)} For each protein, the sum of the charges of all its residues and only the interacting residues on the surface is computed. For each complex, these total and interacting charges from the two interacting partners are multiplied. The bar plot shows, for the whole dataset and each class, the percentage of complexes whose total (in orange) and interacting (in blue) products are negative.
\textbf{d)} The relative abundances of each of the twenty natural amino acids considering all the residues (orange), only the interacting ones (in green), and only the solvent-exposed residues (brown) are shown. The results are divided into four classes. }
\label{Figure1}
\end{center}
\end{figure}

Since electrostatic interactions can act both at short and long distances \cite{mccoy1997electrostatic}, the interaction region that should be considered is of non-trivial definition. 
In fact, smaller regions are able to capture the binding properties due to van der Waals forces, which lead to a shape complementarity between the two interacting molecular surfaces. On the other hand, larger regions lose shape complementarity but involve more charged residues, which are typically excluded from the binding regions but are widely present in the other exposed regions, playing a crucial role in the thermal stability of the protein structure \cite{Thermometer_2022}. It is in fact known that electrostatic interactions between 9 and 12 \AA~are of crucial importance to distinguish obligate from non-obligate complexes \cite{maleki2013role}, thus playing a key role in the characterization of the binding. 
For a quantitative description of the Lennard-Jones potential at the interfaces of protein complexes, we have recently shown that shape complementarity is maximized for interacting patches less than 9 \AA~\cite{Milanetti2021}, highlighting thus the effect of van der Waals interactions at the binding interface \cite{Desantis2022}. 
For a patch to the interface of this size, where the involvement of long-range electrostatic interactions is reduced, the quantification of the contribution of electrostatic complementarity remains unclear.

Here, we characterize the role of electrostatic interaction in protein binding and quantitatively measure the electrostatic complementarity at the interface of the molecular complexes, defining the binding interfaces in a 9 \AA~radius sphere. 
With this aim, we collect a dataset of protein complexes (see Methods for details) and characterize the complexes in terms of interface type, amino acid composition, and charge properties. To study the relationship between binding affinity and electrostatic complementarity we consider a second dataset, 'Affinity' (see Methods for details).
Next, we analyze the contribution to the binding of electrostatic, by comparing the potential values of mirroring points on binding regions. We show that assessing only the concordance (or discordance) of the potentials does not distinguish between interacting and random regions, whereas taking into account the actual values results in higher accuracy. 
Finally, we show how it is possible to extend the 2D Zernike method, that we proposed to quickly evaluate the shape complementarity at interfaces \cite{Milanetti2021, Milanetti2021Insilico,latto,bo2020exploring,biom11121905, Grassmann2022}, to distinguish interacting and noninteracting patches by describing their electrostatic potential projections with a vector and looking at the difference between these descriptors.
Here, for the first time, we describe a computational strategy to compactly describe the electrostatic potential in a single vector composed of Zernike coefficients, where both positive and negative values of the electrostatic potential are considered together in the same function.
For what concerns shape complementarity, our method has already been demonstrated to be able to efficiently identify interacting regions by measuring the shape complementarity in terms of the Euclidean distance between the Zernike invariant descriptors associated with the projections of the molecular surfaces patches (see Methods for more details).
As a final step, we show that the Zernike descriptions of the electrostatic allow for fast and superposition-free discrimination between transient and permanent interactions.

\section{Results}

\subsection{Charges distribution and compatibility}

\begin{figure}[] 
\begin{center}
\includegraphics[width=\textwidth]{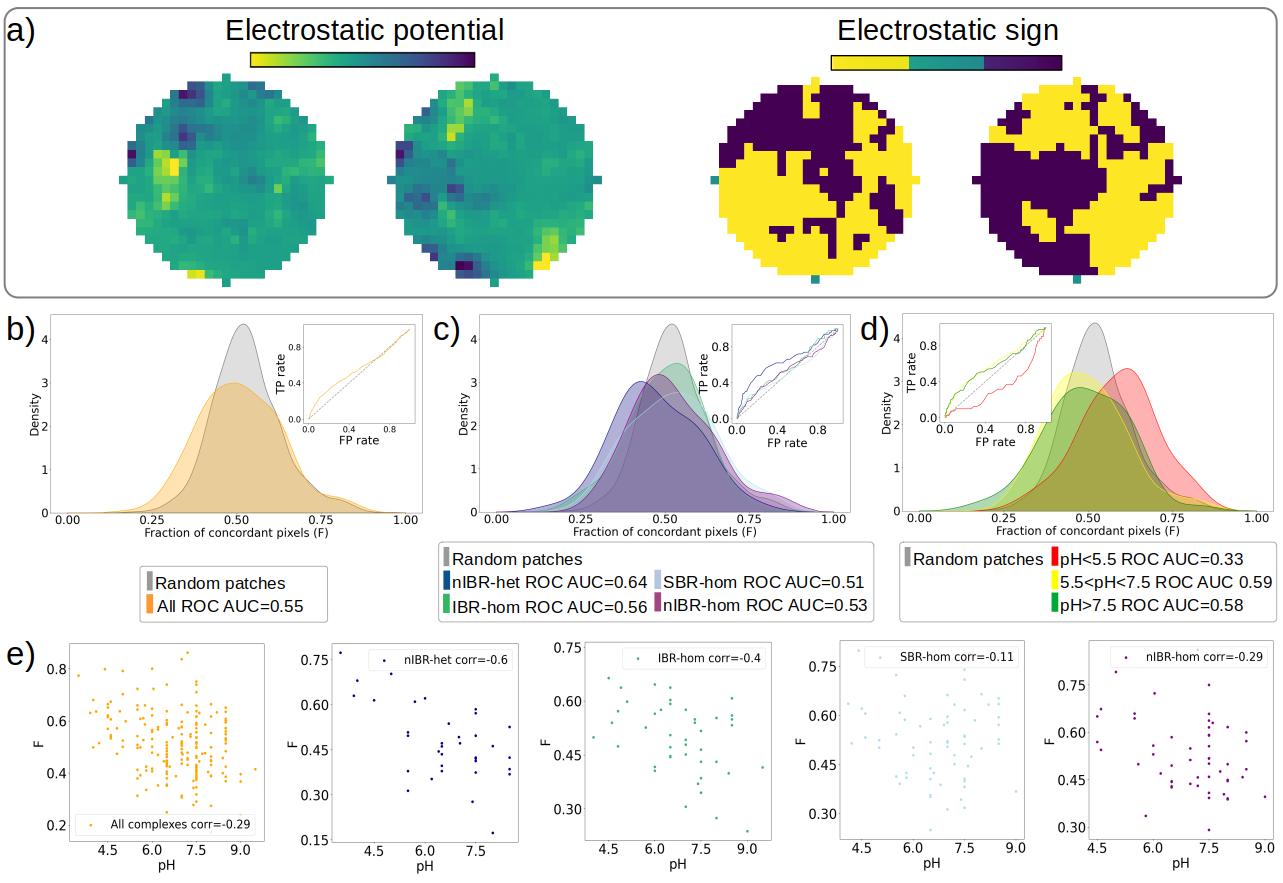}
\caption{\textbf{Electrostatic complementarity contribution in protein-protein complexes} \textbf{a)} On the left, the $EM$s of two interacting patches. Each pixel of the matrices is colored according to the electrostatic potential value of the surface points projected in that region. 
On the right, the $SEM$s of the same two patches: yellow, violet, and cyan pixels correspond respectively to positive, negative, or null values of the electrostatic potential.
 \textbf{b)} Distributions of the $F$ values computed for interacting (orange) and random (grey) patches taken from the 'Human' dataset. In the insert the corresponding ROC curve.
\textbf{c)} The distribution of the $F$ values of the interacting patches in b) is divided according to the class of the complex: nIBR-het in violet, IBR-hom in green, SBR-hom in light blue and nIBR-hom in pink. In the insert the corresponding ROC curves.
\textbf{d)} The distributions of the $F$ values computed for interacting patches in b) are classified in pH ranges: low in red, physiological in yellow, and high in green. In the insert the corresponding ROC curves.
\textbf{e)} Fraction of concordant regions as a function of the pH and computed correlation (in the legend). From left to right the considered complexes are the whole 'Human' dataset, the nIBR-het, the IBR-hom, the SBR-hom, and the nIBR-hom.
 }
\label{Figure2}
\end{center}
\end{figure}

To characterize the role of electrostatic complementarity in protein binding, we collected a balanced dataset of human protein complexes for which structural data were available (`Human' dataset). The dataset is composed of 164 homodimers, which can be divided into 44 dimers with an Identical Binding Region (IBR-hom), 66 Shifted Binding Region (SBR-hom), and 54 non-Identical Binding Region (nIBR-hom), depending on the similarity of the interacting patches (see Figure~\ref{Figure2}a and Methods section for details); finally, the dataset includes 35 heterodimers (nIBR-het).

To begin with, we investigated the amino acid composition of the proteins in the `Human' dataset and their charge distribution. Figure \ref{Figure1}d) shows a general overview of the amino acid abundances, computed considering all the residues in a protein, only the solvent-exposed ones, and only the ones included in the binding regions (see Methods for details about the definition of solvent-exposed residues and binding regions).
The analysis of the amino acid composition confirms the well-known result \cite{Desantis2022} that hydrophobic amino acids, such as Ile or Met, are uncommon in the solvent-exposed surface of proteins. However, when one of them is present in the exposed regions, it is more likely to find it in a binding site rather than on the rest of the surface. On the contrary, charged amino acids, such as Lys or Glu, are more present on the surface, but the fraction taking part in the binding is relatively small.
Figure~\ref{Figure1}d also shows that IBR homodimers tend to have less charged amino acid on the interacting regions, compared to heterodimers and nIBR homodimers.

Figure \ref{Figure1}b shows a general overview of the fraction of residues with the negative, positive or null charge that can be found near a residue on the interacting patch of the binding partner. The results are divided according to the charge of the interacting residue and for the four classes of the dataset.
For all the classes, and each of the considered radius R defining the surrounding of a residue, only $\sim1.5$\% of the negative interacting residues are close to other negative residues on the other protein in the complex.
Positive residues instead can be in proximity more frequently, between 1.6\% and 3.2\% of the time, depending on the complex category and on R. This condition is particularly common for nIBR homodimers, even more, common than the closeness of oppositely charged residues. 
Oppositely charged residues can be rarely found (0.15-0.25\% of the time) among the interacting patches of SBR homodimers as well.
On the other hand, heterodimers and IBR homodimers have a higher percentage, up to 3.6\% and 3.2\% respectively, of opposite charges facing each other in binding regions. 
Figure \ref{Figure1}b characterizes the residues surrounding non-charged amino acids as well. It can be seen that in general all complexes tend to have non-charges facing each other, confirming the predominantly hydrophobic nature of the interacting regions of the protein-protein complexes \cite{yan2008characterization}.
From these preliminary analyses, we would expect heterodimers and IBR homodimers to be less difficult to characterize in terms of electrostatic complementarity than SBR and nIBR homodimers. This is reinforced by the analysis shown in Figure \ref{Figure2}c), where the percentage of complexes that have a total sum of the charges (considering all the residues of each protein or only the interacting ones) with opposite signs is displayed. According to this analysis, heterodimers are the only class whose interacting patches ($\sim4$\%) have a discordant sum of the charges among binding partners. 

\subsection{The specificity of electrostatic complementarity depends on both  complex kinds and environmental factors}
\label{sec:specificity}

Next, we moved to analyze the spatial disposition of the electrostatic interactions. To do so,  we evaluated the electrostatic potential generated by the protein charges on the molecular surface~\cite{Jurrus2017}. Indeed,  the molecular surface is a high-level representation of the external protein structure; such representation allows for an efficient evaluation of both the geometrical and chemical complementarity. In fact, 
the amino acids composition analysis provides information only about the chemical and physical properties of the residues belonging to the two binding sites, while the resolution of the Poisson-Boltzmann equation provides information on the electrostatic potential distribution at the interfaces, which are due to all the atoms of the protein, even those not directly belonging to the binding site.

To analyze the contribution of electrostatic to the binding between two proteins, we define a quantitative measure of electrostatic complementarity able to differentiate between interacting and non-interacting regions. We start by analyzing the 'Human' dataset and comparing the electrostatic potential values of the binding regions with those of non-interacting regions.
To measure the complementarity,  we describe each patch with a simplified Simplified Electrostatic Matrix ($SEM$), computed starting from the Electrostatic Matrix ($EM$), obtained by projecting the considered region of the electrostatic potential surface on the x-y plane, which is defined as the best fit of the surface points belonging to the specific patch. The matrix is then built on the plane and each pixel is associated with the average value of the electrostatic potential of the points inside of it (see Figure~\ref{Figure2}a and Methods section for more details). 
To obtain the $SEM$, we assign +1, -1, and 0 to all the positive, negative, and null pixels respectively. Figure \ref{Figure2}a) shows on the right the $SEM$s of two interacting patches, and on the left the corresponding $EM$s.
To evaluate the electrostatic complementarity between two patches, we compare each pixel of the $SEM$ describing the first patch with the pixel at the same position on the $SEM$ describing the second one. In this way, we can check if surface points that face each other on the binding partners have electrostatic potentials with opposite signs.
We then define $F$ as the fraction of pixel pairs in which the two pixels have the same sign: patches with a high electrostatic complementarity, i.e. a high number of opposite neighbor points with discordant electrostatic potentials, should have a low value of $F$.
The measured complementarity is compared with that one would obtain by chance, randomly selecting surface points and building around each one a patch on the surface of the partner protein.

Figure \ref{Figure2}b shows the distribution of $F$ scores for the whole 'Human' dataset.
In particular, one can see that random patches (i.e. decoys) have a gaussian-like distribution with a mode of 0.49, as we expect that for two random patches the probability that a spatial corresponding region has an opposite sign is 0.5.   
The distribution of all complexes instead is shifted toward values lower than 0.5. In particular, it has a mode of 0.38, indicating that protein binding regions have a degree of electrostatic complementarity higher than what one would expect by chance.
Indeed, this can be quantified by computing the ROC curve (see inset Figure~\ref{Figure2}b) and evaluating the Area Under the Curve (AUC), which is 0.55 in this case.

Looking at the shape of the distribution, one can see that it appears to be composed 
of different populations of proteins; in fact, it presents bi/tri modalities.  
We thus proceeded to separate the dataset according to the complex classes.

Indeed, we found different behaviors for the various classes. 
To quantify the differences between the distributions, we evaluated (see Figure \ref{Figure2}c the ROC curves of each distribution with respect to the decoy's one and computed the corresponding AUC. 
The classes with the lowest AUC of the ROC curve are SBR and nIBR homodimers  (at 0.51 and 0.53 respectively), for which the classification performance can not be distinguished from that of random decoys. A slightly better classification is obtained for the IBR homodimers, with an AUC of the ROC curve of 0.56.
On the other hand, the complexes whose binding regions are more easily classified are the heterodimers, with an AUC of the ROC curve of 0.64.

Interestingly,  a trend is observed also stratifying the dataset according to the pH of each complex. Figure \ref{Figure2}d shows the distributions and relative ROC curves for three ranges of pH values, i.e. high (pH$>7.5$), low (pH$<5.5$), and physiological ($5.5<$pH$<7.5$). Complexes in the high range have a high degree of electrostatic complementarity, having an AUC of the ROC curve of 0.58, whereas the $F$ score of interacting patches in the low pH range is shifted to higher values (resulting in an AUC of the ROC curve of 0.33), meaning that in this case, the binding regions have a higher fraction of concordant points facing each other.

Moreover, in Figure~\ref{Figure2}e, we show the fraction of concordant regions as a function of the experimental pH value for every complex of the `Human' dataset and each subclass. It is interesting to note that while the whole 'Human' dataset does not show a strong correlation with the pH (-0.29), the correlation values vastly differ among the subclasses. The $F$ values computed for SBR and nIBR homodimers are randomly distributed, with a correlation of -0.11 and -0.29 respectively. On the other hand, IBR homodimers and heterodimers anti-correlate with the pH value, correlating -0.4 and -0.6.
Nevertheless, Table \ref{Table1} shows that independently from the class, the interacting patches of complexes in the low range have an AUC lower than 0.5, meaning that for low pH the electrostatic potential in points facing each other on interacting patches has the same sign. Table \ref{Table1} reports as well how the performances of the $F$ score for increasing radius R of the patch: after $R=12 $\AA~ its characterization of interacting patches does not improve or even worsen. Even if larger regions include more charged residues by extending out of the hydrophobic binding sites, the complementarity of the charges is lost when the surfaces of the complex are not interacting. This analysis confirms our choice of a 9 \AA~ radius to define the interacting patches.

\subsection{Low-affinity interactions use electrostatic complementarity to achieve specificity}
\label{sec:stability}

Since both the stratification by classes and pH did not fully account for the observed shape of the distribution, we look for binding affinity data. To do so, we collected a dataset of complexes with known structure and experimental dissociation constant, $K_d$. In particular, we took the dataset proposed by Desantis~\textit{et al.}, which we refer to as the 'Affinity'  dataset  (see Methods for details), which is exclusively composed of 123 heterodimers. 

\begin{figure}[]
\begin{center}
\includegraphics[width=\textwidth]{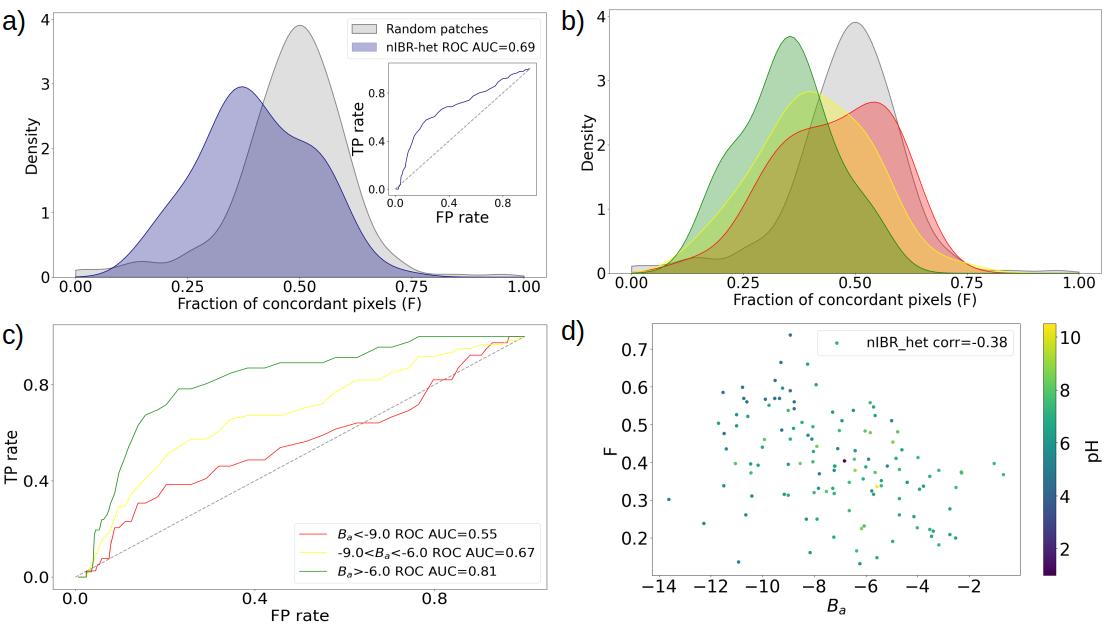}
\caption{\textbf{Electrostatic complementarity contribution in the binding stability of complexes.}
\textbf{a)} Distributions of the $F$ values computed for
interacting (violet) and random (grey) patches taken from the ’Affinity’ dataset. In the insert the corresponding ROC curve.
\textbf{b)} The distribution of the $F$ values of the interacting patches in a) is divided according to the binding affinity of the complex: high affinity in red, medium in yellow, and low in green. In grey is the distribution of the random decoys.
\textbf{c)} ROC curves and corresponding AUC (in the legend) of the distributions of the interacting patches in b), computed against the random distribution.
\textbf{d)} Fraction of concordant regions as a function of the binding affinity and computed correlation (in the legend). Each point is colored according to the pH value, as indicated by the color bar.}
\label{Figure3}
\end{center}
\end{figure}

Figure~\ref{Figure3}a shows the $F$ score distribution of interacting and random patches for the whole 'Affinity' dataset. The latter have a gaussian-like distribution with a mode of 0.49, as for the random patches of the 'Human' dataset. The former instead is shifted to lower values, having a mode of 0.31 and an AUC of the ROC curve of 0.69 (as shown in the insert of Figure \ref{Figure3}a.


Stratifying the dataset in three groups according to the complex affinities, we obtain the results shown in Figure ~\ref{Figure3}b-c. The three distributions, corresponding to high ($B_a<-9.0$), medium ($-9.0<B_a<-6.0$) and low ($B_a>-6.0$) binding affinity are well separated and shifted on different ranges of $F$ scores. 
Low-affinity complexes are moved to lower values of $F$, resulting in an AUC of the ROC curve of 0.81, while the ones with high binding affinity can not be distinguished from random decoys, having an AUC of the ROC curve of 0.55. The medium binding affinity complexes cover an intermediate range of $F$ values and have an AUC of 0.67.
Interestingly, if we look at the $F$ value of each complex as a function of its binding affinity, we get a negative correlation ( -0.38)  with the binding affinity.

\begin{table}
\centering
\begin{tabular}{l|llll}
\textbf{F} & $R=6$ & $R=9$ & $R=12$ & $R=15$  \\
\hline
$pH<5.5$ & 0.32 & 0.33 & 0.33 & 0.35 \\ 
$5.5<pH<7.5$ & 0.54 & 0.59 & 0.61 & 0.62 \\  
$pH>7.5$ & 0.52 & 0.58 & 0.62 & 0.60 \\ 
\end{tabular}
\caption{\textbf{AUC of the ROC curves of the $F$ score for varying patch radius and pH.} The AUC of the ROC curves are computed using the random distribution and the distributions of the 'Human' dataset interacting patches, divided according to the pH. Increasing values of the radius R defining the patches are tested.
}
\label{Table1}
\end{table}

\begin{table}
\begin{tabular}{l|llll}
\textbf{F} & $R=6$ & $R=9$ & $R=12$ & $R=15$  \\
\hline
$B_a<-9.0$  & 0.54 & 0.55 & 0.51 & 0.54 \\  
$-5.5<B_a<-9.0$ & 0.67 & 0.68 & 0.67 & 0.68 \\ $B_a>-5.5$ & 0.8  & 0.85 & 0.88 & 0.87 \\ 
\end{tabular}
\caption{\textbf{AUC of the ROC curves of $F$ for varying patch radius and binding affinity.} In the left table, the AUC of the ROC curves is computed using the binding and random regions $F$s distributions. The binding region distributions are divided into three groups, according to the binding affinity $B_a$ of their complex. Increasing values of the radius R defining the patches are tested. The complexes are part of the 'Affinity' dataset.}
\label{Table2}
\end{table}

\subsection{ Projecting the molecular surface on an orthogonal basis allows to compactly describe the electrostatic contribution to the interface of complexes}

Leveraging on the results of the previous sections, we looked for a compact method to simultaneously measure both electrostatic and shape complementarity between protein patches. 

To describe and compare the electrostatic surfaces more efficiently, we apply the 2D Zernike polynomials, which constitute a complete basis in which any function of two variables defined in a unitary disk can be decomposed. The Zernike expansion associates each portion of a surface to an ordered set of numerical descriptors, invariant under rotation, allowing an easy and fast metric comparison between different protein regions for complementarity evaluation (see Methods for details). The rotational invariance is a fundamental property in the blind search for interacting patches. The complementarity of the binding regions can then be evaluated in terms of the euclidean distance between their corresponding Zernike vectors. In particular, we measure how much the distance between the Zernike descriptors of a pair of interacting sites is smaller than the distances between random patches.

\begin{figure}[] 
\begin{center}
\includegraphics[width=0.8\textwidth]{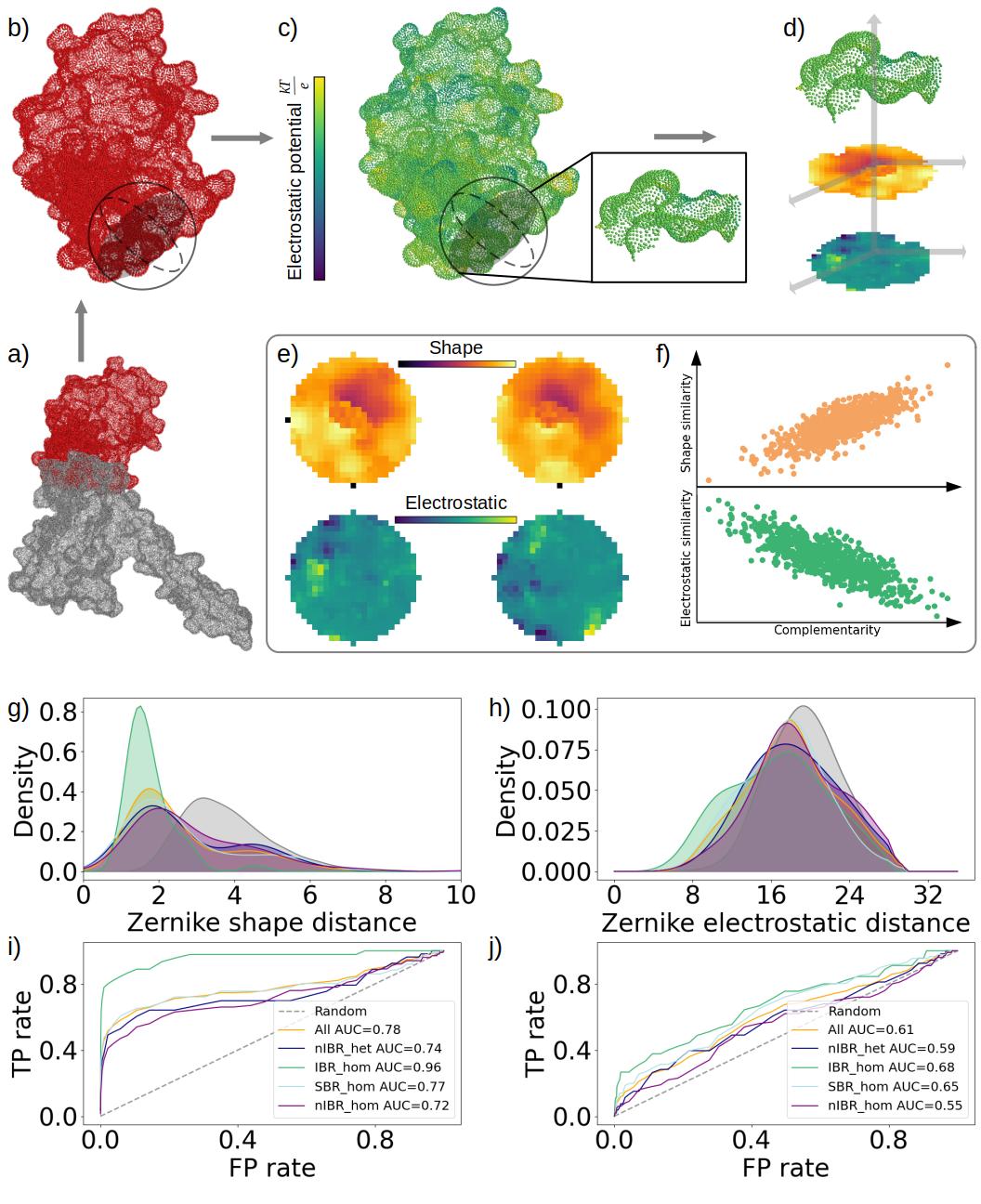}
\caption{\textbf{Schematic representation of the computational protocol.} \textbf{a)} Molecular representations of the surfaces of two proteins forming a complex. \textbf{b)} Molecular representation of the surface of one of the proteins depicted in a). A sphere is used to select a possible patch on the surface: the dark shadow highlights the selected points. \textbf{c)} Electrostatic potential surface, where each point is colored according to its electrostatic potential value. In the zoom, the region of the electrostatic potential surface corresponding to the patch selected in b).\textbf{d)} 2D projections of the patch. In the orange scale the shape projection, for which the colors in the plane are determined by the distance of the surface points from a predefined origin (see Methods for details). In the green scale the electrostatic projection, where the colors are determined by the electrostatic potential values of the above points. \textbf{e)} Comparison between the shape and electrostatic projections of two binding regions. \textbf{f)} Simulated characterization of the complementarity between two interacting patches in terms of shape and electrostatic similarity (top and bottom respectively).
\textbf{g)} Distributions of the distances between the Zernike vectors describing the molecular surface of interacting (orange) and random (grey) patches in the 'Human' dataset. The distribution of the interacting regions is also divided according to the class of the complexes: nIBR-het in blue, IBR-hom in green, SBR-hom in light blue and nIBR-hom in purple.
\textbf{h)} For each patch the distance between the Zernike vectors describing the electrostatic potential surface in that region is computed. Then the same analysis and classification as in g) is performed.
\textbf{i)} ROC curves of the distributions in g) and corresponding AUC (in the legend) computed against the random distribution.
\textbf{j)} ROC curves of the distributions in h) and corresponding AUC (in the legend) computed against the random distribution.
}
\label{Figure4}
\end{center}
\end{figure}

Figure \ref{Figure4} shows a schematic representation of the computational protocol for comparing, in terms of shape and electrostatic, interacting proteins. For each protein, the molecular surface and the electrostatic potential surface are built. The former corresponds to the solvent-accessible surface, the latter is obtained by assigning to each point of the molecular surface the value of the electrostatic potential computed in that region as obtained by solving the Poisson-Boltzmann equation \cite{Chakavorty2016}.
On each surface, a patch is iteratively selected, and the corresponding regions of both the molecular and electrostatic surfaces are separately projected onto a plane. More details can be found in the Methods. 

As a first step we asses the shape complementarity between the interacting patches, by expanding in terms of Zernike polynomials the 2D projections of the molecular surfaces, an example of which is shown in Figure \ref{Figure4}e. As shown in Figure \ref{Figure4}g our results are in line with what has been observed in \cite{Milanetti2021}: interacting patches are efficiently distinguished from random decoys, with an AUC of the ROC curve of $\sim$0.8, as shown in the inset of Figure \ref{Figure4}i. 
The class whose interacting sites can be better identified includes IBR homodimers (with a success rate of 0.96), whereas the lowest efficiency is obtained for nIBR homodimers (AUC at 0.72).
Next, we implement the Zernike method for the study of electrostatic complementarity. 
Since Zernike coefficients can represent only real-valued functions over the unit disk we define, starting from each $EM$, the Confined Electrostatic Matrix ($CEM$). Examples are shown in Figure \ref{Figure4}e.
$CEM$s are obtained by capping the $EM$s pixels above +30 and below $-30$ at those values \cite{Gainza2019}. This allows us to rescale all values so as to obtain Zernike-expandable functions.
We now define electrostatic complementarity as the distance between the Zernike vectors associated with the $CEM$s.
Figure \ref{Figure4}j shows that this definition of complementarity reaches an efficiency in distinguishing interacting and random patches comparable with the one obtained with $F$ values, with an AUC of the ROC of 0.61.
IBR and nIBR homodimers correspond to the best (0.68) and worst (0.55) performances.
As previously assessed, complexes with low pH have a low electrostatic complementarity and can not be easily distinguished from random decoys, reaching an AUC of 0.55, as shown in Table \ref{Table2}. Interestingly, the opposite behavior can be observed for the shape complementarity, which is higher when the pH is low.
\begin{table}
\begin{tabular}{l|ll}
 &  $Zernike_{Shape}$ & $Zernike_{Electrostatic}$  \\
\hline
$pH<5.5$  & 0.84 & 0.55  \\  
$5.5<pH<7.5$ & 0.76 & 0.63  \\  
$pH>7.5$ & 0.77  & 0.61  \\ 
\end{tabular}
\caption{\textbf{Discriminating power of the Zernike-based expansion of molecular and electrostatic potential surface for different pH ranges.} In the $Zernike_{Shape}$ column, the AUC of the ROC curves is computed from the distribution of the Zernike distances between the molecular surface of interacting patches and the random decoys. In the $Zernike_{Electrostatic}$ column the AUC of the ROC curves is computed considering the electrostatic potential description of the patches. The results are divided according to the pH of the complexes.}
\label{Table2}
\end{table}



\begin{figure}[] 
\begin{center}
\includegraphics[width=0.8\textwidth]{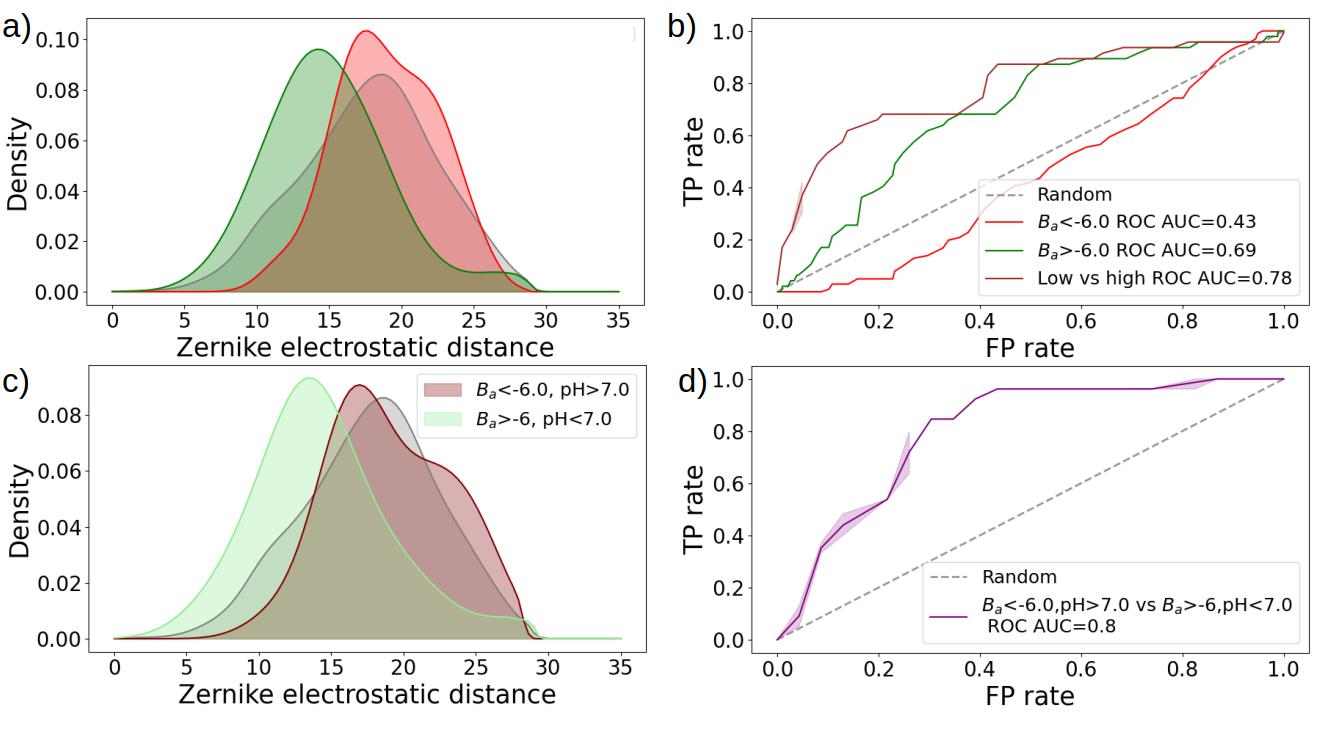}
\caption{\textbf{Superposition-free classification of transient and permanent interactions.} \textbf{a)} Probability density functions of the Zernike electrostatic distances of the 'Affinity' dataset' complexes. Green  (respectively red) distribution corresponds to complexes having $B_a$ values higher (respectively lower) than -6, corresponding to transient and permanent interactions, respectively. Grey curves correspond to the distances of random decoy patches on the protein surfaces. \textbf{b)} ROC curves of the transient (green) and permanent (red) interactions with respect to the decoy distribution, together with the ROC curve of the transient distribution with respect to the permanent one. \textbf{c)} Same as in a) but considering transient complexes with pH higher than 7 and permanent interactions with pH lower than 7. \textbf{d)} ROC curves of the two distributions displayed in panel c.}
\label{Figure5}
\end{center}
\end{figure}

\subsection{Transient from permanent interactions can be distinguished solely based on the electrostatic complementarity}

At last, we apply our method to the 'Affinity' dataset to test the ability of our descriptor to distinguish between permanent and transient interactions, as this property has important effects on biological function \cite{nooren2003diversity}. In particular, defining permanent (respectively transient) interactions based on the binding affinity~\cite{La_2013} being lower (resp. higher) than $B_a = -6$, we obtained the distributions shown in Figure~\ref{Figure5}a. Interestingly, transient interactions display higher than random electrostatic complementarity values (green distribution), while permanent interactions (red distribution) have Zernike distances slightly higher than that one would expect by chance. This can be quantified again by looking at the ROC curves in Figure~\ref{Figure5}b and evaluating the AUC values. Indeed, transient interactions display an AUC of 0.69 with respect to the decoy distribution, while permanent interactions have an AUC of 0.44. Permanent interactions can be distinguished from transient ones with an AUC of the ROC of 0.78. 
Notably, knowing the pH of the considered complexes allows for an even better classification as one can see from Figure~\ref{Figure5}c-d.

\section{Discussion}
The full mapping of the organisms' interactomes is fundamental for understanding molecular interactions and their many physiological and pathological implications. The well-tested toolbox of experimental techniques we dispose of, such as X-ray crystallography \cite{Maveyraud2020-wn}, NMR \cite{takahashi2000novel, foster1998chemical} and cryo-EM \cite{bai2015cryo,cheng2015single},  is allowing for the detection of protein-protein binding and the determination of the complexes atomistic structure. However, all these techniques are expensive and time-consuming \cite{berggaard2007methods} so that up to now only small fractions of the organisms' interactomes have been experimentally determined at the structural level \cite{gu2011prin,plewczynski2009interactome,li2004map}.
In this respect, computational methods represent a powerful tool to unveil the uncharted landscape of protein complexes \cite{zhang2012structure} by predicting protein-protein associations in normal conditions and under mutations/modifications \cite{ezkurdia2009progress, lichtarge1996evolutionary,wang2006predicting,brender2015predicting}, which further complicate the compilation of the interactomes by increasing the number of matches to probe.

To predict the protein complex,  
the identification of putative binding interfaces plays a key role and most of the proposed strategies identify the interfaces as those showing some geometrical/chemical complementary between the molecular partners.
In particular, the side chain rearrangements minimizing the van der Waals interaction determine shape complementarity at the interfaces, which is typically evaluated by geometrical approaches requiring structural alignment between the two interacting molecules.  

Notably, the geometric complementarity of the final complexes does not depend on the dynamical specifics of the binding process. In fact, partners can undergo very few changes upon binding (i.e they follow a “lock-and-key” model) or the interactions between two approaching structures can induce conformational changes (“induced fit”) or the protein conformation suitable for binding (bound state) can be explored by the protein even in the absence of the molecular partner (the “conformational selection” model); these three views suggest different key contributors to the conformational changes between the unbound and bound structures, but for all of them shape complementarity is a necessary condition for the complex formation.
Usually, by including the electrostatic contribution to the binding process investigation, in addition to the Van der Waals forces, one aims at more precise discrimination of the biological interfaces.

In this respect, computational methods can be divided into two categories: model-based and feature-based approaches. The former exploits the residue-conservation found between similar proteins, the latter is based on local features of protein sequences and/or structures.
Feature-based approaches are more general and can work on any type of protein. Even if the availability of protein structures is less abundant than sequences, structural features are fundamental to understanding binding between proteins. Moreover, the recent advances in the field of protein structure prediction starting only from its amino acid sequence \cite{Varadi2021}, vouch for an even more important role of the structural-based method than in the past years. 
Nonetheless, even using structural information, the identification of interfaces remains a challenge in structural biology.
Machine learning-based approaches, give promising results, but they require the definition and training of several parameters and lack a clear physical–chemical interpretation.

Analysis of the electrostatic potential of protein-protein complexes has led to the general assertion that protein-protein interfaces display electrostatic complementarity. Even though a well-settled definition of how electrostatic complementarity should be quantified is still missing, many studies have been highlighting its importance at the interfaces of biological complexes \cite{Bauer2019}.  


Here, we probed the role of electrostatic complementarity in the identification of binding regions and complexes' stability. To do so, we collected two large datasets of protein dimers with known structural information stratified by dimer type and stability (quantified by means of experimental binding affinity). At first, we analyzed the amino acid composition of the binding region with respect to those of the proteins' cores and solvent-exposed regions. 

Next, we looked at the presence and disposition of the charged residues on the binding regions finding that different classes of dimers have slightly different disposition/abundances of charged-charged interactions. Finally, we further increase the complexity of the electrostatic description,  considering the full electrostatic potential generated by the protein partial charges on the solvent-exposed molecular surface. This representation allows for a high-level measurement of the electrostatic complementary at the interface of the interacting molecules. Indeed, comparing both the spatial correspondence of the potential sign (see F descriptor), we found that the binding regions exhibit a complementary higher than the one we could expect by chance. Notably, the signal is influenced both by the complex class and the experimental pH and binding affinities.    In particular, we observed that the maximum complementarity is shown by homodimers that bind using the same region, while homodimers not sharing common residues in the binding region exhibit a nearly random match. 
Notably, the same analyses produce qualitatively identical results if carried out considering the match of the full electrostatic potential, even if they are overall noisier.  
Finally,  we propose a novel method to assess electrostatic complementarity without the need of having complex structures. 
Indeed, we already developed a novel computational protocol based on the Zernike polynomials to describe the shape of portions of the molecular surface in the form of a vector of numbers  \cite{Milanetti2021, Milanetti2021Insilico,latto,bo2020exploring,biom11121905}.  Here, the method is extended to molecular surfaces for which the electrostatic potential has been calculated through the Poisson-Boltzmann equation \cite{Chakavorty2016}. Indeed, after a proper projection of the electrostatic potential surfaces on the 2D plane, electrostatic complementarity can then be defined again as the Euclidean distance between these new Zernike invariant descriptors (see Methods).

In conclusion,  we found that electrostatic complementarity in the binding region is efficiently measured simply requiring a spatial match between the signs of the electrostatic potentials. Moreover, such complementarity strongly depends on both the kind of the considered complex, the pH of the environment, and the transient/permanent nature of the binding. Our results thus help shed light on the often contrasting conclusions of previous works that measured electrostatic complementarity using large datasets. 
Leveraging on our findings, we adapted the Zernike formalism to measure both shape and electrostatic complementarity in a fast and superposition-free manner. 


Finally, we note that our findings could be used to reinforce the docking algorithm, or/and to perform pose selection. Moreover, our method could be adapted to other properties that can be described with numerical values assigned to each surface point, since the Zernike expansion can be applied to any function.


\section{Methods}

\subsection{Protein complex datasets}
To probe the degree of electrostatic complementarity in protein-protein binding regions, we collect a  dataset of protein-protein dimers for which structure information was available from the 3D complex database~\cite{Levy_2006}.   
Selecting only non-redundant human dimers, with an x-ray crystal resolution better than 3.0 \AA~ and no missing residues in the binding region, we ended up with 199 human protein complexes in  PDB format~\cite{Berman2000}.
We opted to restrict to only one organism to avoid spurious effects on the charges distribution in the protein structure, due for instance to thermal adaptation~\cite{Miotto_2018, diffusion, Desantis2022}.

Looking at the dimer composition and spatial orientation, we classify the dataset into four groups:
\begin{itemize}
    \item 44 homodimers with Identical Binding Regions (IBR-hom), i.e. binding regions that have at least 70\% of common residues.
    \item 66 homodimers with Shifted Binding Regions (SBR-hom), i.e. interacting patches that have between 30\% and 70\% of common residues.
    \item 54 homodimers with non-Identical Binding Regions (nIBR-hom), i.e. binding regions that share less than 30\% of the residues.
    \item 35 heterodimers (nIBR-het), where two different proteins are interacting.
\end{itemize}
To analyze the correlation between electrostatic complementarity and binding stability we consider a second dataset, composed of 123 complexes extracted from the dataset used in \cite{Desantis2022} and
with experimental data of binding affinity $B_a$, defined as the $log_{10}$ of the equilibrium dissociation constant $K_d$ \cite{vangone2015contacts}.

\subsection{Computation of the surfaces and surface residues definition}

The solvent-accessible surface for each structure of the dataset was computed using DMS \cite{Richards1977AreasVP}, with a density of 5 points per $\AA^2$ and a water probe radius of 1.4 $\AA$. 
For each surface point, the unit normal vector was calculated with the flag $-$n.
Starting from these surfaces, the electrostatic potential of each protein was calculated independently from the partner using the APBS code \cite{Jurrus2017}, considering the experimental pH. 
The electrostatic potential surface was then defined by building a grid and selecting the values of the electrostatic potential in the grid cells corresponding to each surface point.\\
To select among the residues included in the surface the mainly superficial ones, we computed the Relative Solvent Accessibility as the ratio between Solvent Accessibility and the maximum Solvent Accessible Surface Area of the considered amino acid.
The Solvent Accessibility is calculated with DMS by computing the portion exposed to the solvent of each residue involved in the interaction, while the maximum Solvent Accessible Surface Area of the twenty natural amino acids was taken from \cite{Tien2013}.
A residue is considered superficial if it has a Relative Solvent Accessibility higher than 0.25. 
The interacting regions were defined as the points on a protein surface closer than 6\AA~ to its partner surface.

\subsection{Patch definition and projection}
To define a surface patch, we use a spherical region with radius $R_s$ centered at one point of the surface. This point is randomly extracted for the decoy random patches, while to study the binding regions the geometrical center of the experimental interacting regions is considered. For this study, we chose $R_s=9$\AA~ to be able to study simultaneously the shape and electrostatic complementarity with the 2D Zernike-based method. Indeed, in previous work, we discussed the range of $R_s$ values resulting in the best identification of binding regions when considering shape complementarity \cite{Milanetti2021}.\\
Once the patch has been selected, we re-orient the coordinates.
When two random patches are compared, for each patch we build a plane passing through it and we orient the coordinates so that the z-axis
is perpendicular to the plane. It must be remembered that when comparing the shape of patches, their relative orientation must be evaluated: to assess their shape complementarity, we have to orient the patches contrariwise, i.e. one patch with the solvent-exposed part toward the positive z-axis (‘up’) and one toward the negative z-axis (‘down’).\\ 
On the other hand, to compare the $EM$s and $SEM$s of interacting patches we compute the mean of the normal vector of the first partner and the inverse of the normal vector of the second one. The binding patches are then rotated so that this averaged vector is along the z-axis. This step results again in two patches contrariwise oriented, but in addition to this, we can preserve the spatial correspondence of the surface points after the rotation.
We want to remark here how this correspondence is not necessary when the projections are decomposed in the Zernike basis and compared with the Zernike protocol, giving the rotation invariance of the Zernike polynomials. Therefore, to study the patches in terms of the Zernike polynomials expansion we reorient each binding site along the z-axis independently from its partner.\\
Once the patches have been rotated, two protocols can be implemented. The first one is used to obtain the projections of the corresponding regions of the electrostatic potential surface, whereas the second one provides the projections of the molecular surfaces.

\subsubsection{Electrostatic projection:}
Each point of the re-oriented electrostatic surface is projected on the XY plane. Next, we build a square grid (15$\times$15 pixels) and associate each pixel with the mean value of the electrostatic potential
of the points projected inside of it, and call it the Electrostatic Matrix ($EM$).

\subsubsection{Shape projection:}
Once the patch has been rotated, given a point C on the z-axis we define the angle $\theta$ as the largest angle between the z-axis and a secant connecting C to any point of the patch. C is then set so that $\theta=45^\circ$.\\
To study the shape of the patch, each surface point is labeled with its distance r to C. We then build a square grid (15$\times$15 pixels), associating each pixel with the mean r value calculated on the points inside it.

\subsection{Zernike 2D protocol}
Each function of two variables $f(r,\psi)$ defined in polar coordinates inside the region of the unitary circle ($r<1$) can be decomposed in the Zernike basis as
\begin{equation}
f(r,\psi)=\sum_{n'=0}^\infty\sum_{m=0}^{n'}c_{n'm}Z_{n'm}(r,\psi),
\end{equation}
where
\begin{equation}
    c_{n'm}=\frac{n'+1}{\pi}\int_0^1dr~r\int_0^{2\pi}d\psi Z_{n'm}^*(r,\psi)f(r,\psi)
\end{equation}
and
\begin{equation}
    Z_{n'm}=R_{n'm}(r)e^{im\psi}.
\end{equation}
$c_{n'm}$ are the expansion coefficients, while the complex functions $Z_{n'm}(r,\psi)$ are the Zernike polynomials. The radial part $R_{n'm}$ is given by
\begin{equation}
    R_{n'm}(r)=\sum_{k=0}^{\frac{n'-m}{2}}\frac{(-1)^k(n'-k)!}{k!\big(\frac{n'+m}{2}-k\big)!\big(\frac{n'-m}{2}-k\big)!}.
\end{equation}
Since for each couple of polynomials, it is true that
\begin{equation}
   \braket{ Z_{n'm}|Z_{n''m'}}=\frac{\pi}{n'+1}\delta_{n'n''}\delta_{mm'},
\end{equation}
the complete sets of polynomials form a basis, and knowing the set of complex coefficients ${c_{n'm}}$ allows for a univocal reconstruction of the original patch. The resolution of this reconstruction depends on the order of expansion $N=max(n^{'}).$\\
The norm of the coefficients $z_{n'm}=|c_{n'm}|$ define the Zernike invariant descriptor, which is invariant for rotations around the origin of the unitary circle.\\
The complementarity between two given patches defined with a sphere of radius $R_s$ can then be measured as the Euclidean distance between the two corresponding invariant vectors: the more the complementary the smaller the distance between their corresponding Zernike vectors. This evaluation can be applied to any properties of the patches that can be described by assigning a numerical value to each surface point.\\
The efficiency of this method depends on two key parameters: the radius $R_s$ and the Zernike maximum expansion order $n$.  When $R_s$ is too low, the patches lack sufficient surface to distinguish the compatibility between interacting regions, whereas too-large patches would include non-interacting regions that have a low complementarity per se. $n$, on the other hand, determines the level of details captured: too low orders could confuse interacting and random patches because the surfaces are excessively “smoothed”, while an excessively accurate level of description would model unnecessary (and time-consuming) details.\\
In this study, we performed the Zernike protocol using $R_s=9$\AA~ and $n=20$, in accordance with the most efficient parameters identified in the previously mentioned work \cite{Milanetti2021}.

\section*{Acknowledgment}
The research leading to these results has been also supported by European Research Council  Synergy grant ASTRA (n. 855923).



<


\end{document}